# Reversible doping of graphene field effect transistors by molecular hydrogen: the role of the metal/graphene interface

C. L. Pereira[1,2,*], A. R. Cadore[1,2,*], N. P. Rezende[1], A. Gadelha[1], E. A. Soares[1], H. Chacham[1], L. C. Campos[1], R. G. Lacerda[1,*]

[1]*Departamento de Física, Universidade Federal de Minas Gerais, Belo Horizonte, 30123-970, Brasil*

[2]*These authors contributed equally to this work.*

*)*Electronic mail: cintialpfisica@gmail.com; alissoncadore@gmail.com, rlacerda@fisica.ufmg.br*

In this work, we present an investigation regarding how and why molecular hydrogen changes the electronic properties of graphene field effect transistors. We demonstrate that interaction with $H_2$ leads to local doping of graphene near of the graphene-contact heterojunction. We also show that such interaction is strongly dependent on the characteristics of the metal-graphene interface. By changing the type metal in the contact, we observe that Ohmic contacts can be strongly or weakly electrostatically coupled with graphene. For strongly coupled contacts, the signature of the charge transfer effect promoted by the contacts results on an asymmetric ambipolar conduction, and such asymmetry can be tunable under interaction with $H_2$. On the other hand, for contacts weakly coupled with graphene, the hydrogen interaction has a more profound effect. In such situation, the devices show a second charge neutrality point in graphene transistor transfer curves (a double-peak response) upon $H_2$ exposure. We propose that this double-peak phenomenon arises from the decoupling of the work function of graphene and that of the metallic electrodes induced by the $H_2$ molecules. We also show that the gas-induced modifications at the metal-graphene interface can be exploited to create a controlled graphene p-n junction, with considerable electron transfer to graphene layer and significant variation in the graphene resistance. These effects can pave the way for a suitable metallic contact engineering providing great potential for the application of such devices as gas sensors.

## 1 – INTRODUCTION

Graphene is a two-dimensional (2D) material that is well known by its remarkable thermomechanical and electronic properties. The ability to modulate and control graphene electronic properties in different and efficient ways are of paramount importance for the commercial realization of graphene applications [1–4]. For instance, graphene field effect transistors (GFETs) have demonstrated great potential for applications in several areas such as high-frequency transistors [2,5], biological and chemical sensors [6,7], and in a variety of optoelectronic devices [8–11]. In the GFET configuration, an electric current between the source and drain terminals is modulated by the application of a gate potential [1,5,12,13], In this context, it is well known in the microelectronics industry that electric transport between the conducting channel and metallic electrodes plays an important role in the device characteristics and performance [14,15]. Thus, an intensive effort has been devoted to understand and improve the electrical junction between the metallic electrodes and graphene [16–22]. Typically, in a metal-semiconductor junction, there is a Schottky-barrier that actively affects the electronic device properties [3,5]. However, in a metal-graphene junction, even an Ohmic contact can passively play a role in device conduction due to unusual effects originated from the quantum electronic properties of graphene [23]. Moreover, graphene depicts a zero-gap and approximately linear energy dispersion near the vicinity of the Fermi level [4,24]. These features provide an ambipolar electronic conduction, and GFET transfer characteristics (conductance versus gate voltage curves) typically displaying a "V-shaped" form [1]. Thus, the minimum of conductivity (or maximum of resistivity) of the transfer curve corresponds to the crossing of the Fermi level at the Dirac point in the graphene electronic band structure [4]. Around this point, where the graphene density of states vanishes, the concentration of thermoactivated electrons is equal to that of holes being labeled as the charge neutrality point (CNP) [1,4]. In general, there is only one CNP in the transfer characteristics of GFETs [1,2]. Nevertheless, several experimental works have demonstrated that graphene devices can exhibit multiple maxima of resistance depending on external environmental conditions, contaminations, and device structuring [25-31]. This means that, in such systems, there are adjacent regions of graphene with different local density of charges, or even the existence of heterojunctions (regions p-type doped nearby regions n-type doped) as it is theoretically predicted [15]. A double-CNP in the transfer curve of a GFET can also be intentionally created using a double-gate structure that enables setting the density of charges at different regions of the graphene channel independently [34,35]. Apart from this case, all the other processes that create the double-CNP are irreversible, which limits the device electrical performance and applicability. Therefore a further understanding of these phenomena, as well as the use of it in a controllable and tunable way is still lacking.

In this work, we systematically investigate how and why molecular hydrogen tunes the electrical properties of graphene devices. We show that the interaction with $H_2$ promotes a local doping of graphene at the metal-graphene interface, which is strongly dependent on the characteristics of the device. For instance, by changing the type of metal-graphene heterojunction, we observe that Ohmic contacts can be strongly or weakly coupled electrostatically with graphene. In a strongly coupled heterojunction, there is a Fermi level pinning effect at the metallic leads fixing the charge density nearby. Consequently, graphene field effect transistors show an asymmetric ambipolar conduction, and such asymmetry can be tunable under interaction with $H_2$ [23]. On the other hand, in devices with metallic leads weakly coupled with graphene (such as Au/$Cr_2O_3$ electrodes), hydrogen interaction at the metal/graphene interface generates a second charge neutrality point in graphene transfer curves (a double-CNP feature). We propose that this double-peak phenomenon arises from the decoupling of the work function of graphene and that of the metallic electrodes under interaction with $H_2$. The induced double-CNP or "M-shaped" form observed is completely controllable and reversible via molecular hydrogen ($H_2$) exposure. Those effects provide considerable electron transfer to graphene layer and large variation in the graphene resistance providing an alternative approach for suitable engineering at graphene/contact interfaces with great potential for application of gas sensors.

## 2 - EXPERIMENTAL METHODS

For a better understanding of the role of hydrogen on the electronic properties of GFET, we built several devices with different parameters: device geometry, channel lengths, and types of metallic electrodes. Graphene devices were prepared using monolayer graphene produced via mechanical exfoliation on $SiO_2$(300nm)/Si substrate, where heavily doped silicon issed as the back-gate electrode [1]. Electron beam lithography and oxygen plasma etching were used to define the graphene shape. The metallic electrodes of all devices were designed by electron beam lithography, followed by thermal metal deposition and lift-off. Here, we fabricated GFETs with pure Au (30nm), Au/Cr (30nm/1nm) and Au/$Cr_2O_3$ (30nm/~1nm) as metal-type electrodes, forming the metal-type/graphene/$SiO_2$/Si structures. GFETs were also fabricated with 5nm and 10nm $Cr_2O_3$ thick for the Au/$Cr_2O_3$ transistors. In all these devices, the metallic electrodes were defined and deposited on top of the graphene channel, as illustrated in the top panels of Figures 1(a) and 1(c)-1(f). However, GFETs were also fabricated by transferring graphene on top of pre-prepared Au/Cr (30nm/1nm) electrodes, as illustrated in the top panel of figure 1(b), forming the graphene/Au/Cr/$SiO_2$/Si structures.

Our standard GFETs devices were prepared with several materials as contacts, and keeping the same graphene channel length ($L = 1$µm), channel width ($W = 3$µm), and electrodes length ($d = 1$µm). In the inset of figure 2(b) we illustrate the definition of electrode length, $d$. Additional devices were prepared to investigate the effects of the contact length, graphene channel length, and contact resistance on the formation of secondary CNPs. Accordingly, GFETs with Au/$Cr_2O_3$ electrodes, which show tunable secondary CNPs, were prepared at several other configurations: (i) GFETs with several electrodes lengths (from $d = 0.25$µm up to $d = 2$µm, fixing $L = 1$µm and $W = 3$µm – see inset in figure 2(b)); (ii) GFETs with several graphene channel lengths (from $L = 1$µm up to $L = 10$µm, keeping $d = 1$µm and $W = 3$µm – see inset in figure 2(c));

(iii) GFETs in Hall bar geometry, where we can ignore the influence of contact resistance[20,23]. In this geometry, we also measure the electronic properties of the device in a non-invasive (figure 1(e)) and invasive (figure 1(f)) configuration [20,23]. Finally, the formation of the $Cr_2O_3$ instead of Cr as a sticking layer underneath the gold contacts was produced by intentional oxidation of the chromium after deposition by thermal evaporation. For this process, immediately after the chromium deposition (~ 1 nm), usually performed at pressures around $2 \times 10^{-6}$Torr, the thermal evaporation chamber was opened to the environment for 30min and then pumped down back to $2 \times 10^{-6}$Torr before the gold deposition. Such an oxidation process of Cr was tested and characterized before we implemented it as our device preparation recipe. The oxidized materials (~1 nm of $Cr_2O_3$) were verified by atomic force microscopy (AFM) and X-ray photoelectron spectroscopy (XPS) (see Supporting Information).

After fabrication, each device was inserted into a homemade gas system tube equipped with heater and mass flow controllers, which enables the control of the temperature, from $T = 25°C$ up to T= 200°C, gases flow (Argon (Ar) and $H_2$) with hydrogen concentrations ($[H_2]$), from $[H_2] = 0.5\%$ up to $[H_2] = 50\%$ inside the chamber at atmospheric pressure. Before carrying out the electrical measurements, each device undergoes to a thermal conditioning process that consists on keep the devices at a temperature of $T = 200°C$ under a flow of 300sccm of ultrapure Ar for 8h. This process is known to remove absorbed water molecules, impurities and contaminating gases from the graphene surface [36–38]. For the electrical characterization, we performed mainly two-probe measurements using the standard lock-in techniques, keeping fixed the frequency at 17Hz and current bias ($I_{SD}$) between the source (S) and drain (D) terminals of $I_{SD} = 1\mu A$, while we measured the voltage drop (V) in between both electrodes as illustrated in the top panel of figure 1(a). However, for the Hall bar geometry, the same parameters were used in between source and drain, while the voltage was measured in the inner electrodes as illustrated in the top panel of figure 1(e). Finally, for all devices, the total device resistance (R) was obtained by applying Ohm's law.

## 3 - RESULTS AND DISCUSSION

We start by showing typical transfer curves of a GFET and how $H_2$ modifies graphene electrical properties. In figures 1(a)-(d), we present data of two-probe measurements of transfer curves ($R$ x $V_G$) at $T$=200°C of graphene devices with contacts made of Au, Au/Cr and Au/Cr$_2$O$_3$. Note that graphene is contacted with Au electrodes in different ways. In figure 1(a), Au contacts are at the top of graphene, while in figure 1(b), graphene is at the bottom of the Au electrodes. Firstly, we would like to stress that in all devices characterized graphene main channel are n-type doped. This can be seen by the fact that the CNP locates at negative gate potentials. Such charge transfer effect depends mainly on the activation of surfece dangling bonds in the SiO$_2$ by thermal conditioning, which is not controllable, explaining why the CNPs for different devices are not at the same gate voltage [36,39]. Secondly, the graphene conductance is ambipolar around the charge neutrality point, meaning that at the left side of the CNP, the electrical conduction is carried out by positive charge carries (holes) whereas at the right side of the CNP the conduction is executed by negative charge carries (electrons). Moreover, other information can be extracted by asymmetric transfer curves. If the transfer curve shows a larger resistance at the right side of the CNP, graphene at the contact interface is p-type doped promoting a preferential scattering of electrons [17,20,23]. Similary, if the transfer curve shows a larger resistance at the left side of the CNP, the opposite applies, and we can infer that graphene at the contact interface is n-type doped scattering holes more efficiently than electrons. By carefully analyzing figures 1(a)-(d), one can see that there is a preferential scattering of electrons in absence of $H_2$ (all metallic contacts naturally promote a p-type doping of graphene near the leads). This statement is more evident in graphene/Cr/Au devices as shown in the black curve of figure 1(c) [17,20,23]. However, under exposure to $H_2$ (red curves in figures 1(a)-1(c)), there is an inversion of the asymmetry of the transfer curves [23]. In these circumstances, the metallic contacts promote a n-type doping of graphene in the region near the contacts and resistance at the left side of the CNP is larger that at the right, of one compares the same density of charge. Additionally, under interaction with $H_2$, in all devices the CNP shifts towards more negative values of the gate potential, indicating a global electron transfer to the graphene. Previous works have justified similar charge transfer processes via $H_2$ dissociation and subsequent interaction of the atomic hydrogen with the graphene channel, causing the negative charge transfer and leading to chemical and permanent changes on the GFETs [25,40,41]. However, as we will discuss next, we do not observe permanent changes on the graphene electrical properties induced by $H_2$, and the idea of a hydrogenation process can be disregarded.

Now, by comparing the results presented in figures 1(a)-1(d) it is clear that the CNP shifts ($\Delta V_G$) are not the same for all devices. These data indicate that the charge density transferred via interaction with $H_2$ depends on both the metal-type used and on the electrode position on the graphene channel (top or bottom). In this context, some features can be highlighted. Firstly, the largest charge transfer occurs in graphene devices contacted with pure Au (as we will discuss later, the charge transfer is proportional $\Delta V_G$). Secondly, in GFETs with $Au/Cr_2O_3$ electrodes, as shown in Figure 1(d), in addition to the CNP shift, there is a formation of a second "peak" in the $RxV_G$ curves when the device is exposed to $H_2$ (red curve in figure 1(d)), resulting in a "M-shaped" curve. Here, it is important to mention that all data depicted in figures 1(a)-1(d) are taken from graphene devices with the same geometry (graphene channel length, width and electrodes length).

Now, we state that all effects discussed above can be addressed to charging effects at the graphene/metallic contact region. A demonstration that the graphene main channel is not becoming charged directly due to a interaction with $H_2$ is understood in measurements of devices prepared in a Hall bar geometry, as shown in figure 1(e). In such geometry, it is well-known that the contribution from the contact resistances can be disregarded [20,23]. Indeed, we do not observe any significant charging effect under $H_2$ exposure, even using $Au/Cr_2O_3$ electrodes. A similar result is described in a previous work using Au/Cr electrodes [23]. Nevertheless, when the same device is measured in a two-probe configuration, the second CNP appears as well as the electron charge transfer, as it is shown in figure 1(f). We emphasize that the double-CNP occurs in all devices while keeping the current bias very low ($I_{SD}$ = 1µA), suggesting that a second CNP does not appear as a result of charge trapped in the vicinity of the drain at the graphene/SiO$_2$ interface during a current bias stress [42]. Moreover, we also show in the Supporting Information an investigation of the double-CNP by measuring the hysteresis on the GFETs. Such experiments strongly indicate that contributions from trapped charges at the substrate [12,42,43] on the formation of the second CNP and charge transfer can be neglected. In summary, our experiment suggests a strong correlation between the generation of the secondary CNP via charging effects at the metal-graphene interfaces, which also may affect graphene contact resistance.

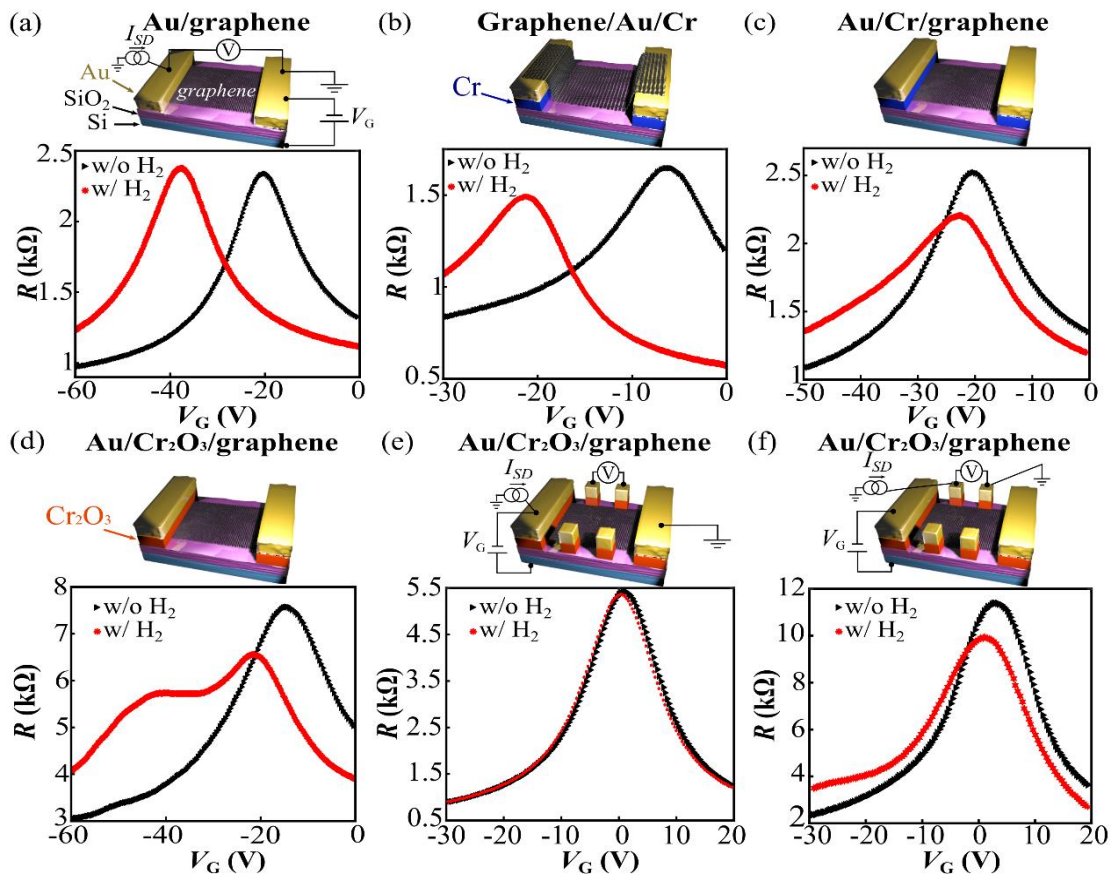

**Figure 1.** Measurements of two-probe resistance ($R$) as a function of back-gate voltage ($V_G$) under H$_2$ exposure for GFETs with different metal-type electrodes: pure Au/graphene (a), graphene/Au/Cr (b), Au/Cr/graphene (c), and Au/Cr$_2$O$_3$/graphene (d) on top of SiO$_2$/Si substrates. All devices are designed with the same graphene geometry ($L$ = 1µm and $W$ = 3µm) and electrodes length ($d$ = 1µm). (e) Four-probe and

(f) two-probe $RxV_G$ curves for a GFET fabricated in Hall bar geometry with Au/Cr$_2$O$_3$ electrodes. All measurements presented in this figure are performed at $T = 200°C$ and the data under *[H$_2$]*=20% exposure (red curves) are performed after 1h of gas exposure, while the black curves are measurements in pure argon, before turning on the molecular hydrogen. The insets in the figures illustrate the metal-type used in the fabrication process, the position of the graphene layer in the GFETs, the device geometry and the configuration for the electrical characterization.

It is important to address possible changes on the contact resistance when the device is exposed to H$_2$. Several works [14,44–46] propose that tunneling across the metal-graphene interface should dominate the charge injection from the metal into the underlying graphene in devices with Ohmic contacts. Besides that, changes on the transmission probability could change the contact behavior from Ohmic to non-Ohmic, also resulting in the observation of multiple CNPs [44–46]. Therefore, we show in figure 2(a) measurements of $I_{SD}xV_{SD}$ for a device with Au/Cr$_2$O$_3$ electrodes before (blue curve) and after (red curve) gas exposure at $T = 25°C$ and *[H$_2$]* = 20%. One can note that the Ohmic behavior (linear relation between $I_{SD}xV_{SD}$) of the device is not affected by the molecules. We only note a change on the slope of the curve shown in figure 2(a), which can be associated with the charge transfer to graphene channel, as demonstrated and discussed in figure 1(d). Such linearity implies that the oxide layer deposited does not form any detectable Schottky barrier in our devices. In addition, under interaction with H$_2$ there is a significant charging effect but the contact region between graphene and Au/Cr$_2$O$_3$ is still Ohmic for any value of gate bias applied (see Supporting Information). Also, in the Supporting Information, we present the $I_{SD}xV_{SD}$ curves for all other devices with different metallic electrodes where a linear response is obtained in all of them (with or without H$_2$ interaction).

Now we investigate how the formation of multiple CNPs under interaction with H$_2$ depends on the device geometry. It has been reported that the two peaks in the $RxV_G$ curves can be associated with the contact area and the length of the graphene channel [47,48]. In such experiments, the nature of contacts, without the presence of any type of gas, might be causing a local charging effect on the graphene underneath the electrodes. So, in figures 2(b) and 2(c) we present curves of the resistivity $\rho$ as a function of the carrier density *n*, for devices with several Au/Cr$_2$O$_3$ electrode lengths (figure 2(b)) and graphene channel lengths (figure 2(c)) after H$_2$ exposure. Here, the resistivity is calculated as $\rho = RW/L$, while the carrier density is obtained by $n = C(V_G - V^{CNP})/e$, where *C* is the graphene device capacitance per unit area for a 300nm SiO$_2$ thick device, $V^{CNP}$ is the back-gate voltage value at the first and more positive CNP, and *e* is the electron charge. A normalization that considers the CNP at more positive voltage values as $n = 0$ cm$^{-2}$ is adopted to better visualize the phenomenon (secondary peaks). As it will be explained later, such CNP presents the electrostatic conditions needed to neutralize charges at the graphene main channel. Figure 2(b) shows the dependence of the second CNP as a function of the electrode length *d*. Clearly, the second CNP is evident in devices with larger metallic electrodes. Such information is a strong evidence that charge is transferred to graphene underneath the contact, and, very likely, such graphene regions acquire a local charge density that is different from that of the graphene main channel. Moreover, both charge carrier type and density are tunable under interaction with H$_2$, corroborating with the hypothesis that the region underneath the contact is important for the interaction with H$_2$ molecules. Note that in this case, the measurements are performed from two neighbor terminals with similar electrodes length *d*, while the graphene geometry is kept fixed at $L = 1$µm and $W = 3$µm. Hence, any contribution from the graphene area can be neglected.

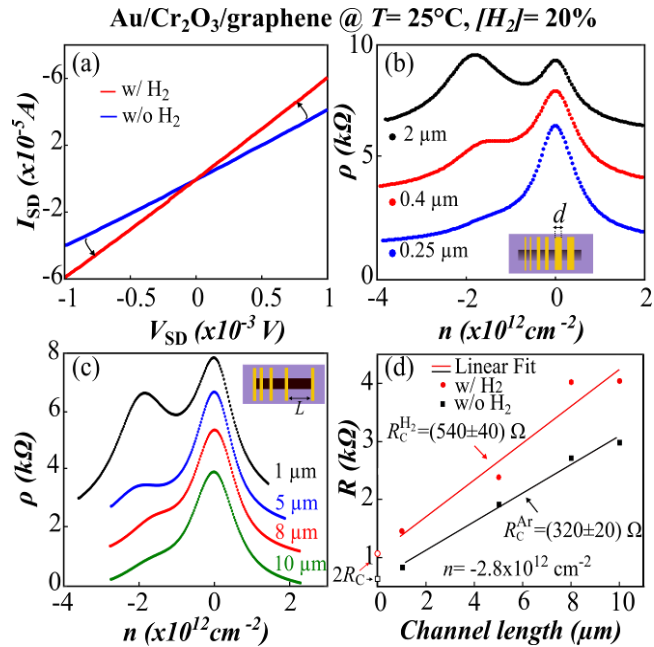

**Figure 2.** (a) $I_{SD} x V_{SD}$ curves for a GFET with Au/Cr$_2$O$_3$ electrodes without (w/o) (blue) and with (w/) (red) H$_2$ exposure. Resistivity ($\rho$) as a function of carrier density (*n*) for different Au/Cr$_2$O$_3$ electrodes length, *d* (b), and graphene channel length, *L* (c) after 1h under H$_2$ exposure. In (a) and (b) the geometry is fixed at $L = 1\mu m$ and $W = 3\mu m$, while in (c) and (d) the electrodes are fabricated with $d = 1\mu m$ and graphene width $W = 3\mu m$. In figures (b) and (c) the curves are shifted to better visualize both CNPs. (d) Transfer length method resistance values measured for different graphene channel distances for the condition without (black) and with (red) gas exposure. All measurements presented in this figure are performed at room temperature ($T = 25°C$) and *[H$_2$] = 20%*.

We also investigate the relationship between the graphene channel *L* and the appearance of the second CNP in GFETs. In figure 2(c) we present data taken from GFETs with contacts of Au/Cr$_2$O$_3$, with several graphene channel lengths, and under *[H$_2$] = 20%* exposure at $T = 25°C$. In all devices, the length of the contacts is kept fixed at $d = 1\mu m$ and the measurements are performed from two neighbor terminals for different graphene lengths. From the $\rho \, x \, n$ curves, we see that secondary CNPs are present mainly in short devices: GFETs with small graphene channel lengths. This data corroborates with the hypothesis that in shorter graphene main channels, the contribution of the graphene region underneath the electrodes are more significant [27,48,49]. Therefore, we can state that the ratio between the graphene and electrodes length is also important for the observation of multiples CNPs, as it would be expected [14,48].

A better presentation of the change of the contact resistance ($R_C$) of the devices before and after $H_2$ exposure is shown in figure 2(d). Here, we present data taken from the same device showed in figure 2(c), at a fixed value of $n = -2.8 \times 10^{12} cm^{-2}$ (left side of the CNP). The total resistance $R$ scales linearly with $L$, and from the linear extrapolation of the resistance, we can estimate the intercept at $L = 0$ µm which gives the sum of the $R_C$ of both electrodes. Average contact resistance is then estimated as half the vertical intercept of the fitting line in Figure 2(d). Such values are presented in black, for devices without interaction with $H_2$ ($R_C^{Ar}$) and, in red, for devices under $H_2$ exposure ($R_C^{H2}$). The fitted $R_C$ values are $R_C^{Ar} = W.(320 \pm 20)\,\Omega\,\mu m$ and $R_C^{H2} = W.(540 \pm 40)\,\Omega\,\mu m$, respectively. Therefore, one can notice that the GFETs exposed to $H_2$ depict higher contact resistance, indicating that $R_C$ is affected by the hydrogen molecules, $R_C^{H2}/R_C^{Ar} \approx 69\%$. We also show in the Supporting Information the analysis of contact resistance for values at the right side of the CNP in figure 2(c). One can note that a similar behavior is observed with a change in $R_C$ and a decrease of the contact resistance of about $R_C^{H2}/R_C^{Ar} \approx 93\%$ after hydrogen exposure. These results suggest that $H_2$ molecules can significantly change the metal-graphene interface potential. For instance, we believe that such molecules modify the p-n junctions at the metal-graphene regions, therefore changing the preferential scattering mechanisms (asymmetry in the transfer curves), as previously discussed. In other words, changes in the doping level at the metal-graphene interface define the preferential scattering for either holes or electrons, modifying then the contact resistance measured. Consequently, this phenomenon explains the increase observed for $R_C^{H2}$ for the left side of CNP (hole-branch) and decrease for the right (electron-branch) side.

Next, we present in more detail the time dependence of the multiple CNPs for the GFETs with $Au/Cr_2O_3$ electrodes. Figure 3(a) shows the evolution of the second CNP as function of the exposure time to $[H_2] = 20\%$ at $T = 200°C$ for a device fabricated with $d = 1µm$. Initially, under an Argon atmosphere (without $H_2$), the $R x V_G$ curve depicts only one CNP at $V_G^{ch} = -7.6V$, which we label it as the CNP position for the graphene channel. However, as soon as there is $H_2$ inside the chamber (adsorption), the electron charge transfer takes place ($V_G^{ch}$ shifts to more negative values $V_G^{ch} = -13.7V$) and a second CNP emerges, stabilizing at around 30min under gas exposure at $V_G^{cont} = -37V$, which we label as the CNP position for the graphene underneath the electrodes. When we turn off the $H_2$ gas (desorption) the system returns to the initial stage, with a single CNP (figure 3(b)). It is important to stress that the interaction of the device with molecular hydrogen is completely reversible. For measurements under $[H_2] = 20\%$ the desorption time is larger than 1.5h. We also need to emphasize that the desorption depends on the length $d$ of the electrodes: decreasing its length also decreases the desorption time. For instance, the desorption time drops from 5h for a length of $d = 2\mu m$ to 0.5h for $d = 0.5\mu m$. This effect is consistent with the diffusion of $H_2$ molecules in between the graphene sheet and the electrodes, indicating that the trapped molecules can easily escape from narrower metallic contacts. Also, the saturation time interestingly decreases for contacts with longer lengths. For $d = 2\mu m$, for example, the saturation time is about 3min, while for $d = 0.5\mu m$ it is 8min. Such non-expected behavior is still not clear to us and this process is under new studies. Manipulation and control over the device time evolution is crucial for providing a faster route to detect molecular hydrogen using metal-type engineering in GFETs.

Even though the data presented in figures 3(a) and 3(b) correspond to a temperature $T = 200°C$, the same behavior occurs for temperatures ranging from $T = 25°C$ up to $T = 200°C$. For instance, figure 3(c) shows $R \: x \: n$ curves for all temperatures analyzed after 1h under $[H_2] = 20\%$ exposure. One can note that the appearance of the second CNP does not depend on the temperature: even at $T = 25°C$ there is the formation of the second CNP. The fact that the second CNP can be observed at room temperature, but with a very small charge transfer process with $H_2$ exposure (see Supplementary Information), would indicate that both phenomena are in fact decoupled from each other, having different origins as we will discuss later. Moreover, from figure 3(c) it is possible to observe an increase in the splitting between the CNPs as the temperature increases, which is also presented in more detail in figure 3(d). In this figure, we show the Fermi energy variation ($\Delta E_{Fermi}$) calculated between both CNPs as a function of temperature. In this case, the Fermi energy variation is defined by $\Delta E_{Fermi} = \hbar v_F \sqrt{\pi |\Delta n|}$, where $\hbar = 6.58 \times 10^{-16} eV.s$ is the Planck's constant ($h/2\pi$), the Fermi velocity is given by $v_F = 1 \times 10^6 m/s$ and $|\Delta n|$ is the difference of the amount of charge density in between both CNPs [31,47]. In this figure, one can note that the splitting in energy increases up to $T = 150°C$ and decreases afterwards for $T = 200°C$. The origin for such decrease is still unclear at the moment. Such results indicate that the interaction at the metal-graphene interface is also favored at higher temperatures, showing that modifications at the metal-graphene interface by $H_2$ molecules are thermally activated. Additionally, the splitting between both CNPs is also dependent on the $H_2$ concentration. In figure 3(e) we present the $R \times n$ curves for all concentrations analyzed (from $[H_2] = 0.5\%$ up to $[H_2] = 50\%$) and after 1h under $H_2$ exposure at $T = 200°C$. This result shows that both the intensity of the second CNP and the splitting between both CNPs depend on the $H_2$ concentration, see figure 3(f). This indicates that the amount of $H_2$ at the interface is determinant for the definition of the doping level at the region underneath the electrodes.

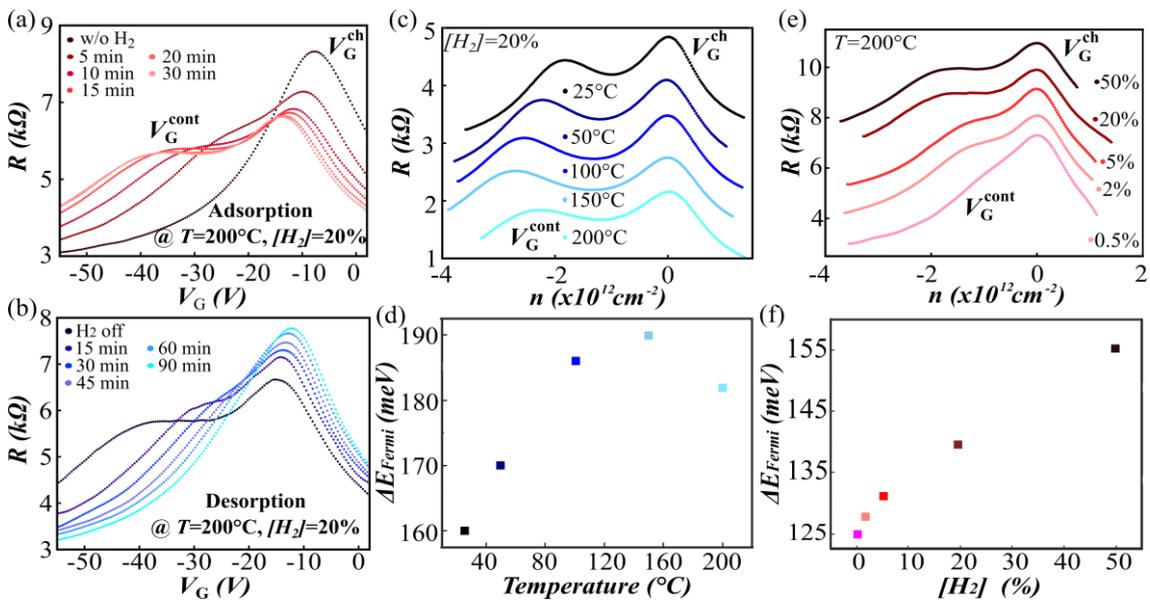

**Figure 3.** $RxV_G$ curves for different time of exposure to $[H_2]=20\%$ in argon during adsorption (a) and desorption (b) processes at $T=200°C$. (c) $R \times n$ for different temperatures, which the measurements are obtained after 1h under exposure to *[H₂]*=20%. (d) Variation of Fermi energy ($\Delta E_{Fermi}$) between both CNPs as a function of temperature. (e) $R \, x \, n$ for different *[H₂]*, where the measurements are obtained after 1h at *T*=200°C. (f) $\Delta E_{Fermi} \, x \, [H_2]$ at *T*=200°C. All the measurements presented in this figure are performed on a GFET fabricated with Au/Cr$_2$O$_3$ electrodes and contact length of *d*=1µm.

Now, it is important to stress that apart from a double-gate structure that enables setting of the density of charges at different regions of the graphene channel independently, all the other processes that create a double-CNP are irreversible [21,28,31,43,47,49]. In our case, the formation of the second CNP is totally controlled and reversible in devices fabricated with a 1nm thick layer of Cr$_2$O$_3$, and under hydrogen exposure. However, if the GFETs are fabricated with 5nm or 10nm of Cr$_2$O$_3$, the second CNP is presented even before H$_2$ gas exposure, but are also tunable under H$_2$ interaction (see the Supporting Information). In such conditions, when the devices are exposed to H$_2$ there is a shift of both CNPs towards more negative gate values, indicating that both regions (graphene main channel and graphene underneath the contacts) accept electrons. We shall also comment that GFETs with thick layers of Cr$_2$O$_3$ are extremely unstable at high temperatures whereas devices fabricated with 1nm Cr$_2$O$_3$ thick are very stable.

We will now address possible explanations for the phenomenology associated to H$_2$ doping. Let us first consider possible physical mechanisms for the n-doping action of H$_2$ molecules on an Au-graphene interface, without the presence of either Cr or Cr$_2$O$_3$, as evidenced by the results of figure 1(a). Our discussion will be based on existing theoretical first-principles results on the literature, as well as on previous experimental evidence. A detailed first-principles study of electronic and structural properties of interfaces between graphene and the 111 surfaces of Al, Ag, Cu, Au, and Pt has been performed in the works of Giovannetti *at al*., and Khomyakov *et al.* [16,50] The first-principles results show that the shift of the CNP of graphene with respect to the Fermi level, Δ*E$_F$*, is an increasing monotonic function of the metal-graphene distance for all interfaces. For the Au-graphene case at the equilibrium distance, the calculation shows graphene as p-type doped, with Δ*E$_F$* of 0.19eV. Therefore, an increase of this distance due to the presence of an $H_2$ interlayer would result in additional p-doping, and not n-doping, from geometrical reasons alone. The calculations also show a positive and monotonic dependence of $\Delta E_F$ with $W_M$-$W_G$, where $W_M$ and $W_G$ are the work functions of the isolated metal surface and graphene, respectively. Among the investigated systems, the ones closest to the neutrality condition ($\Delta E_F = 0$) are Au and Cu, with Au being slightly p-dopant ($\Delta E_F = 0.19$eV) and Cu being slightly n-dopant ($\Delta E_F = -0.17$eV). The calculated work functions $W_M$ of the Au and Cu surface are respectively 5.54eV and 5.22eV. From our measured n-type carrier density of the order of $2\times10^{12}$cm$^{-2}$ under hydrogen exposure, we can estimate, from the electronic density of states of graphene, that $\Delta E_F = -0.16$eV, very similar to the calculated value for Cu. We therefore estimate that the net action of $H_2$ exposure is to reduce the work function $W_M$ of Au by order of 0.3eV and consequently transform the graphene doping from p-type to n-type.

Are there known experiments involving $H_2$ adsorption that could lead to reductions of the Au surface work function $W_M$ by tens of an eV? The answer is "yes" for Au-TiO$_2$-Ti diodes used as high sensitivity $H_2$ sensors [51]. In this work, the Ohmic behavior of the Au-TiO$_2$ junction in hydrogen atmosphere was associated to a reduction of the Au surface work function by several tens of an eV. An additional experimental information on the charge transfer between Au and $H_2$ has been recently reported [52]. These experiments reported a negative charge transfer from Au nanoparticles to adsorbed $H_2$ at ambient conditions, such that these nanoparticles become positively charged. This last information can, in principle, provide a possible mechanism for a negative charge transfer from non-planar regions of Au to graphene in the presence of adsorbed $H_2$.

Now, let us discuss the formation of the second CNP based on the literature. Previous works [17,43,46,49] discussed the observation of multiple CNPs when the Fermi level pinning induced by the electrodes is weak. Hence the graphene underneath the contact can be modulated by gate bias, resulting in changes of the charge density between both regions. Moreover, at the conditions of weak metal-graphene interaction, the conical points in the graphene band structure are preserved, but charge transfer to or from the metallic electrodes can take place, modulating then the Fermi level at the region underneath the contacts [16,50]. Other works also demonstrate that the formation of oxide layer in between graphene and the electrodes could be the reason of multiple CNPs [27,48,49]. In our present case, the second CNP appears with the presence of chromium oxide and under $H_2$ exposure. In an experimental work on chromium oxide gas sensors, Miremadi *et al.*[53] observed the reduction of the conductivity of p-type sensors as a function of hydrogen concentration. This n-type dopant action of $H_2$ was associated with the reduction of $Cr^{4+}$ to $Cr^{3+}$ at the oxide surface. Therefore, we might be facing two distinct mechanisms for $H_2$-induced n-doping, one associated to Au-graphene (as discussed in the preceding paragraphs) and the other associated to chromium oxide-graphene. The latter $H_2$-doping mechanism has been proposed to involve the modification of the oxide surface [53]. We suggest that this mechanism might also result in the decoupling of the electron states of graphene from those of the oxide contact regions. Such decoupling has been considered a necessary ingredient for the appearance of two CNPs in $RxV_G$ curves of GFETs with Ni contacts due to the oxidation of Ni at contact regions [27,48,49], and such $RxV_G$ curves are similar to the one observed in figure 1(d) upon $H_2$ exposure, suggesting a similar physical origin.

Finally, figure 4 illustrates a phenomenological model for the different metal/graphene heterojunctions studied. Figure 4(a) shows that effect of the molecular hydrogen when only Au metallic contacts are used. In this case, the $RxV_G$ curve shows only one CNP and the asymmetry generated by the p-type doping induced by the Au-electrodes and pinning of the graphene work function by the metal [16,17,45]. In this case the Fermi energy ($E_F$) level for the region underneath the contact remains fixed due to the charge density pinning, while the $E_F$ for the channel region can be tunable by the gate bias. However, we suggest that the $H_2$ molecules modify the interface potential by interacting with Au interface reducing its work function and causing the observed n-type doing effect. Now, figure 4(b) depicts the case of GFETs with Au/Cr electrodes without $H_2$ exposure, where the $RxV_G$ curve also shows a single CNP and asymmetry induced by the electrodes and pinning of the graphene work function [16,17,45]. Moreover, similar to the case of pristine gold, the $E_F$ for the region underneath the contact remains fixed, while the $E_F$ for the graphene channel can be tunable by the gate bias. In this case, when the hydrogen is turned on, the $RxV_G$ curve presents a single CNP, and a subtle n-type doping, as illustrated in figure 4(b). Nevertheless, solely for GFETs with Au/Cr$_2$O$_3$ electrodes, the $RxV_G$ curves present two CNPs when the hydrogen is turned on, as we illustrate in figure 4(c). Such observation can be associated to the decoupling of the work functions between the metal and graphene that occurs induced by the presence of H$_2$ molecules at the metal-graphene interface. In the latter case, the $E_F$ of both regions are modulated by the gate bias, and if both regions exhibit different doping levels, the $RxV_G$ curve shows double-CNP, one of them originated from the minimum of density of states in the graphene channel and the other coming from the graphene at the contact region.

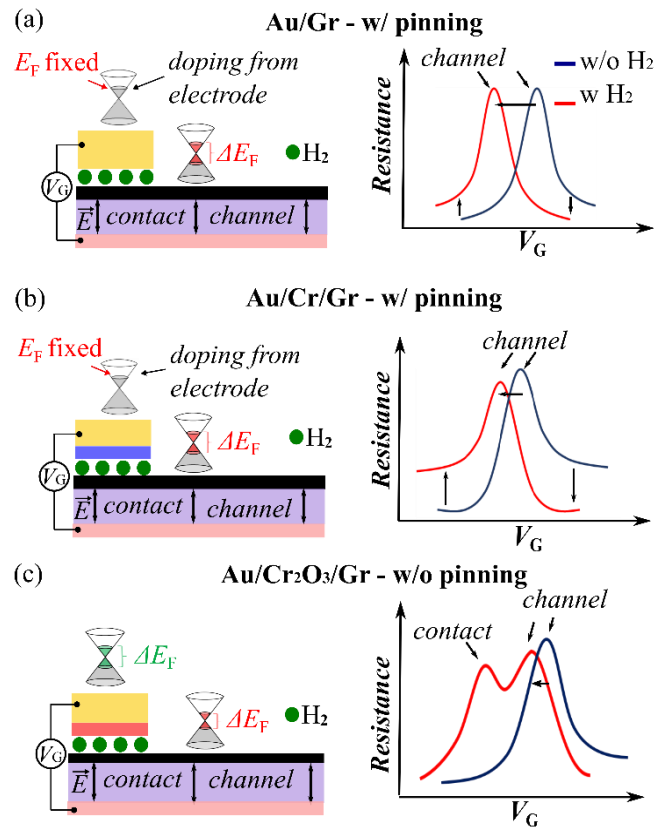

**Figure 4.** (a) A schematic of the graphene device with hydrogen exposure for GFETs with Au metallic electrodes. The band diagram on the contact region indicates the Fermi level pinning ($E_F$ fixed) induced by the metallic electrode, and the band diagram on channel region shows the $E_F$ modulation by the $V_G$ application. The $RxV_G$ curve presents only one CNP associated solely with the graphene channel. (b) A schematic of the graphene device with gas exposure for GFETs with Au/Cr metallic electrodes. The $RxV_G$ curves present also the Fermi level pinning ($E_F$ fixed) before and after $H_2$ exposure. (c) A schematic of the graphene device with gas exposure for GFETs with Au/$Cr_2O_3$ metallic electrodes. The band diagram of the graphene underneath the contact and at the channel regions show the $E_F$ modulation by the gate bias, demonstrating now the existence of double-CNP in the $RxV_G$ curve.

## 4 – CONCLUSION

In summary, we have investigated the formation of multiple charge neutrality points induced in graphene devices by molecular hydrogen exposure. Our findings demonstrate that the observation of the "M-shape" on the $RxV_G$ curves obtained by using thin layers of chromium oxide as metal-type electrodes is not solely dependent on the graphene channel and electrodes size, but also on hydrogen concentration and the temperature at which the interaction occurs.

Moreover, our study confirms that the formation of the second CNP is totally reversible and indicates that after $H_2$ exposure the graphene regions underneath the electrodes can be modulated by the gate bias. Therefore, our results are valuable for two reasons: first, for a better understanding of the metal-graphene interface and the formation of multiple CNPs; second, it shows that the electrode engineering can be used to improve the hydrogen detection using non-functionalized graphene devices.

## ACKNOWLEDGEMENTS

This work was supported by CAPES, Fapemig (Rede 2D), CNPq and INCT/Nanomateriais de Carbono. The authors are thankful to Lab Nano at UFMG for allowing the use of atomic force microscopy and Laboratório de Cristalografia at UFMG for allowing the use of X-ray excited photoelectron spectroscopy. The authors also acknowledge I. S. L. Antoniazzi for the AFM measurements and V. Ornelas for improvements in the gas sensing system employed in this work.

Supporting Information

# Reversible doping of graphene field effect transistors by molecular hydrogen: the role of the metal/graphene interface

C. L. Pereira[1,2], A. R. Cadore[1,2], N. P. Rezende[1], A. Gadelha[1], E. A. Soares[1], H. Chacham[1], L. C. Campos[1], R. G. Lacerda[1,*]

[1]*Departamento de Física, Universidade Federal de Minas Gerais, Belo Horizonte, 30123-970, Brasil*

[2]*These authors contributed equally to this work.*

*)*Electronic mail: rlacerda@fisica.ufmg.br*

**1 – Identification of the chromium oxide layer ($Cr_2O_3$) by Atomic force microscopy (AFM) and X-ray excited photoelectron spectroscopy (XPS)**

The chromium oxide layer is obtained using a thermal evaporation chamber as the following: immediately after chromium (Cr) deposition (~1nm), performed usually at pressures around $2 \times 10^{-6}$ Torr, the thermal evaporator chamber is opened to atmospheric pressure for 30min (to promote the Cr oxidation), then pumped back down to $2 \times 10^{-6}$ Torr, before the gold deposition on the top of it. However, to identify the formation of the oxide layer and its stoichiometry the final top gold layer was not deposited. Then, the oxidized Cr film was characterized by Atomic Force Microscope (AFM) and x-ray excited photoelectron spectroscopy (XPS). Figure S1(a) shows the topography image of the chromium oxide layer thermally evaporated on $SiO_2$/Si obtained by AFM. As can be seen the oxidized Cr layer has a rough-like structure with about 2 nm of thickness. This kind of morphology is expected for very thin layer films deposited by thermal evaporation.

Figure S1(b) shows the XPS analysis of chromium oxide. With this result, we can infer the stoichiometry of the oxide formed under ambient conditions. The long range energy spectrum shows photoelectron peaks that can be assign to Cr, O, C and Si. From the high resolution spectrum in the Cr 2p region (inset of Fig.S1(b)), we determine that the binding energy of the Cr $2p^{3/2}$ peak is 576.4 eV and that the spin orbit splitting of the 2p level is 9.7 eV. The reported values of these quantities in the literature for $Cr_2O_3$ are 576.9 eV and 9.8 eV, respectively [1] .Therefore, we can conclude the formation of a $Cr_2O_3$ layer in our sample.

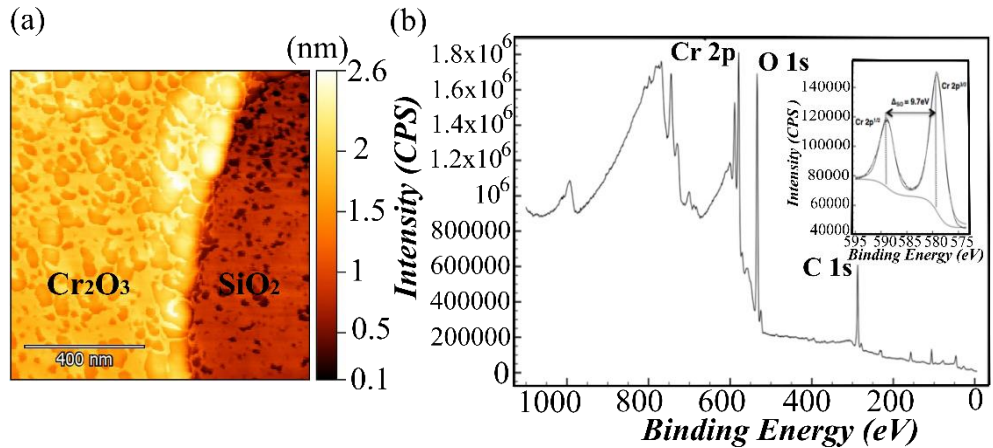

**Figure S1**. (a) Atomic force microscopy of $Cr_2O_3$ film (1nm). (b) X-ray excited photoelectron spectroscopy after the oxidation of 1nm of chromium.

## 2 – Two-probe measurements in devices with Hall bar configuration

As described in the main manuscript text, graphene devices that are fabricated in a Hall bar configuration and exposed to $H_2$ gas do not present neither charge transfer nor the formation of the double-CNP. However, if the same device is measured in a two-probe configuration, both features can be observed. For instance, Figure S3(a) shows the two-probe $RxV_G$ measurements between the more distant contacts (L=15μm), while Figure S3(b) exhibits for the nearest terminals (L = 7μm). In the primer case, the emergence of second CNP under $H_2$ exposure is not evident but becomes clear for the shorter channel where a second CNP can be seen at $V_G^{ch} = -25V$. Note that such results are in total agreement with Figure 2(d) in the main text, where we described the dependence of channel length with the observation of two CNPs.

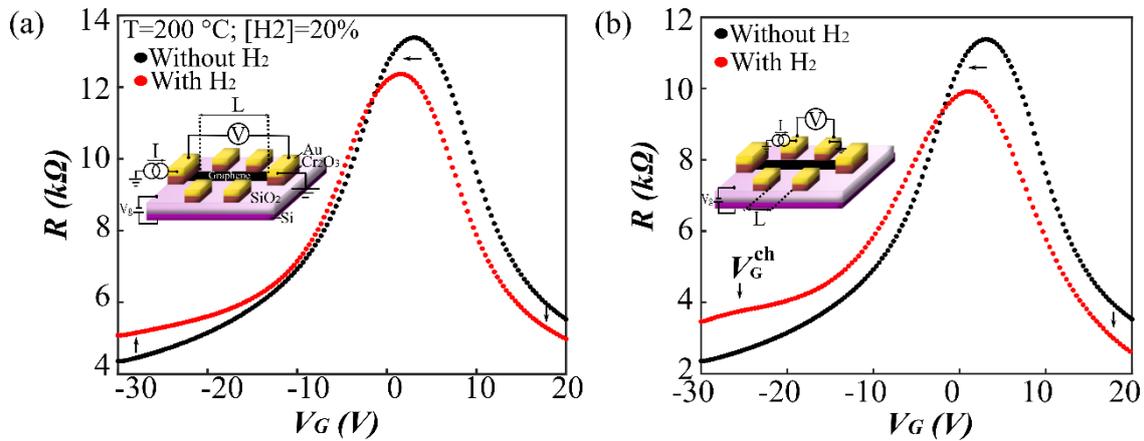

**Figure S3**. Two-probe $RxV_G$ curves for a GFET fabricated in Hall bar geometry with Au/Cr$_2$O$_3$ electrodes between the (a) farthest terminals ($L$=15µm) and (b) closest ones ($L$=7µm). Measurements presented in this figure are performed at $T$=200°C and the data under $[H_2]$=20% (red curves) are performed after 1h of gas exposure, while the black curves are measurements in pure argon, before turning on the molecular hydrogen.

### 4- Hysteresis of the GFETs with different metal-type electrodes

Figure S4 presents an investigation of the double-CNP by measuring the hysteresis of the GFETs for all metal-type electrodes. The left panels present the $V_G$ forward scan (indicated by the blue arrows) with (red) and without (black) H$_2$ exposure, while the right panels present the same case for $V_G$ backward scan (indicated by the green arrows). The measurements show that the $RxV_G$ curves before and after exposure to hydrogen for all types of metal electrodes studied in this work have the same shape, regardless of the direction of the sweep of the applied potential window. This means that the observed effects due to hydrogen, both n-type doping and the appearance of the second peak, are not associated with trapped charges at the graphene-substrate interface [2,3].

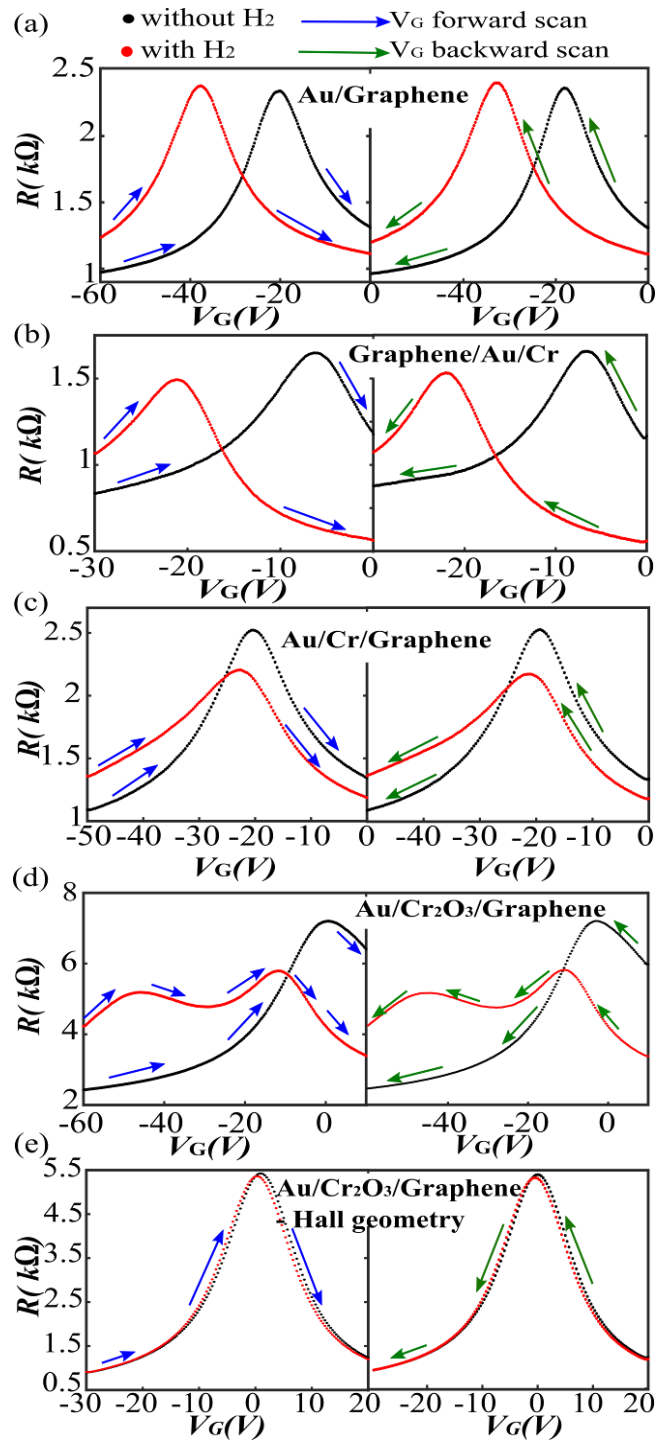

**Figure S4**. $RxV_G$ curves with (red) and without (black) $H_2$ exposure for GFETs with pure Au (a), Cr/Au (b), Au/Cr (c), Au/Cr$_2$O$_3$ (d) electrodes. While in (e) is the GFET in hall geometry with Au/Cr$_2$O$_3$ electrodes.

## 5 – $I_{SD}xV_{SD}$ curves for different configurations of metal-types

Figure S5 shows $I_{SD}xV_{SD}$ curves of all the metal-type configurations used in our work. In all the curves, for each contact geometry (represented in the insets), an Ohmic behavior is observed before (blue) and after (red) exposure to hydrogen. One can note that the linear relation between $I_{SD}xV_{SD}$ of the device is not affected by the molecules, while the small variation in the current values is associated with the charge transfer to graphene channel.

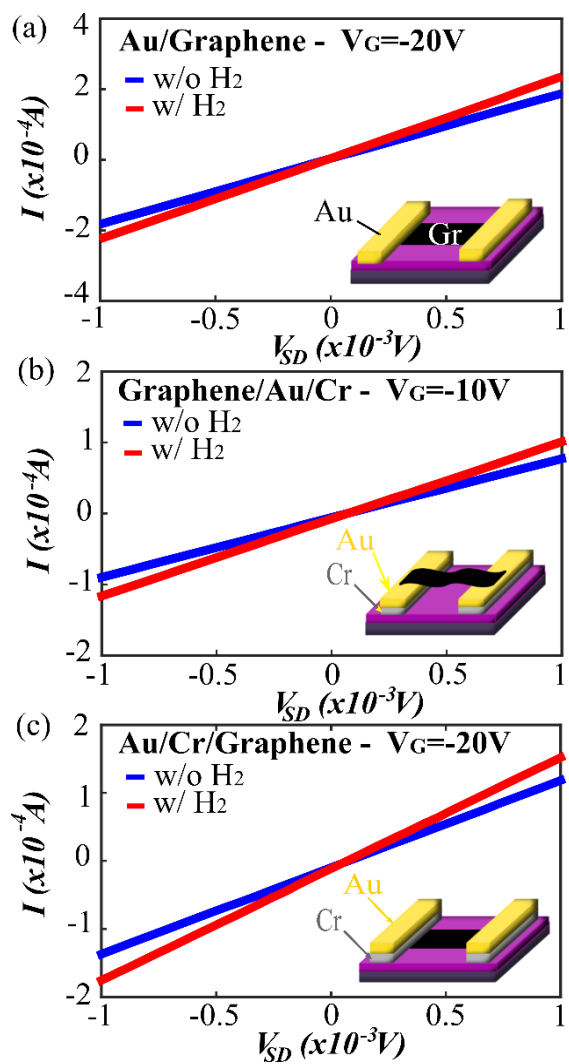

**Figure S5**. $I_{SD}xV_{SD}$ curves without (w/o) (blue) and with (w/) (red) H2 exposure for a GFET with (a) Au, (b) Cr/Au and (c) Au/Cr.

In more detail, Figure S6 shows the $I_{SD}xV_{SD}$ curves as a function of $V_G$ for Au/Cr$_2$O$_3$/graphene devices, where the Ohmic behavior appears for all $V_G$ values. Figure S6(a) shows a single value of $V_G$ in which the current decays and subsequently increases, indicating a single CNP before gas exposure. This aspect can be clearly seen in the $RxV_G$ curve presented in the inset of the same figure. However, Figure S6(b) shows two distinct values of $V_G$ at which the current decreases, indicating the formation of two CNPs after the exposure (see inset of the same figure).

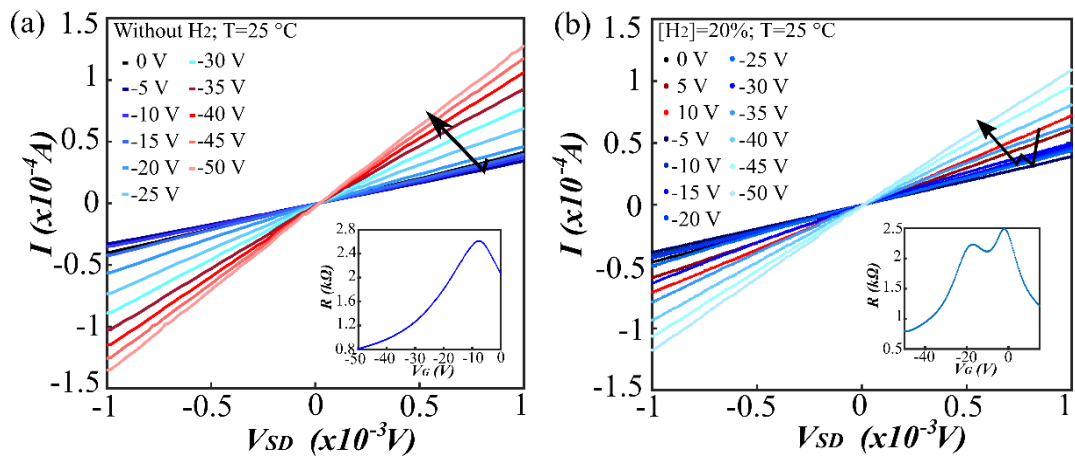

**Figure S6**. (a) $I_{SD}xV_{SD}$ curves w/o H$_2$ exposure for a GFET with Au/Cr$_2$O$_3$ electrodes. The inset shows the two-probe $RxV_G$ curve at the same conditions. (b) $I_{SD}xV_{SD}$ curves under H$_2$ exposure for the same device. The inset brings the $RxV_G$ curve at the same conditions. The measurements were carried out 1h after hydrogen exposure. All the measurements presented in this figure were performed at $T$=25°C and $[H_2]$=20%.

## 6 – Electron charge transfer for different graphene devices at room temperature

Figure S7 shows the $RxV_G$ curves for the same conditions of temperature (25°C) and $[H_2]$=20%, for the different metal-types electrodes used in this work. The $RxV_G$ curves for all devices do not present any significant shift for more negative values of gate voltage. For instance, one can note that the position of the CNP which is associated with the graphene slightly shifts to negative values but no larger than $\Delta V_G^{ch} = 3V$, demonstrating a small charge transfer at room temperature. However, as mentioned on the text, such shift can be as large as $\Delta V_G^{ch} = 20V$ at $T$=200 ºC, demonstrating that the charge transfer is a thermally activated process.

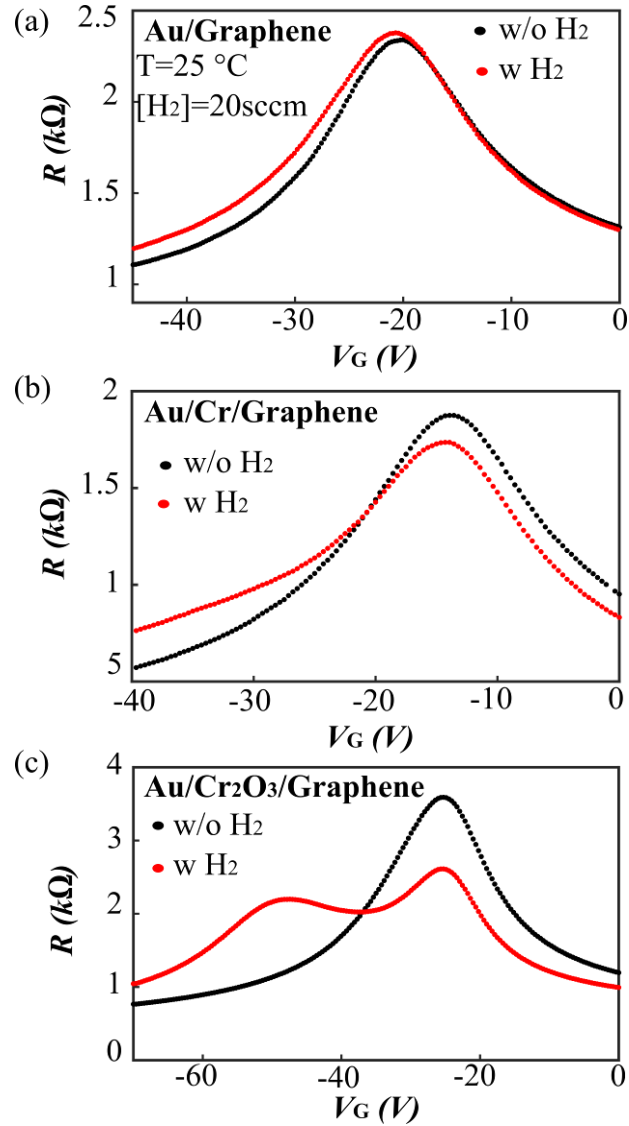

**Figure S7**. $RxV_G$ curves at room temperature (25°C) and $[H_2]$=20%. The devices are prepared with (a) Au electrodes, (b) Au/Cr electrodes and (c) Au/Cr$_2$O$_3$ electrodes.

### 7 – Graphene devices prepared with a different $Cr_2O_3$ thickness

As described in the main text, all the samples measured using 1nm of $Cr_2O_3$ do not show any additional CNP after fabrication and without $H_2$ exposure. However, devices were also fabricated using a thicker oxide layer, such as 5nm and 10nm $Cr_2O_3$ but with a similar top gold layer thickness (30nm), forming the $Au/Cr_2O_3/graphene/SiO_2/Si$ structures. The $RxV_G$ curves of these new devices always present two CNP without any $H_2$ exposure at ambient temperature and Argon atmosphere, as shown in Figure S8. For both $Cr_2O_3$ thickness tested, the curves present immediately the "M-shaped" form, independently of the environment. For such devices is not clear which CNP is related to the graphene channel region or underneath the electrodes. However, we consider that such oxide thicknesses are enough to induce the decoupling between both work functions, as observed in previous works [4,5]. Therefore, comparing the results for thin oxide layers as presented in the main text, we believe that the same behavior is valid. In other words, the CNP positioned at $V_G^{ch} = -22V\ (-29V)$ for the curves present on Figure S8(a) (Figure S8(b)) is associated with the CNP for the graphene channel for 5nm (10nm) oxide thick. While $V_G^{cont} = -37V\ (-45V)$ is related with graphene underneath the electrodes.

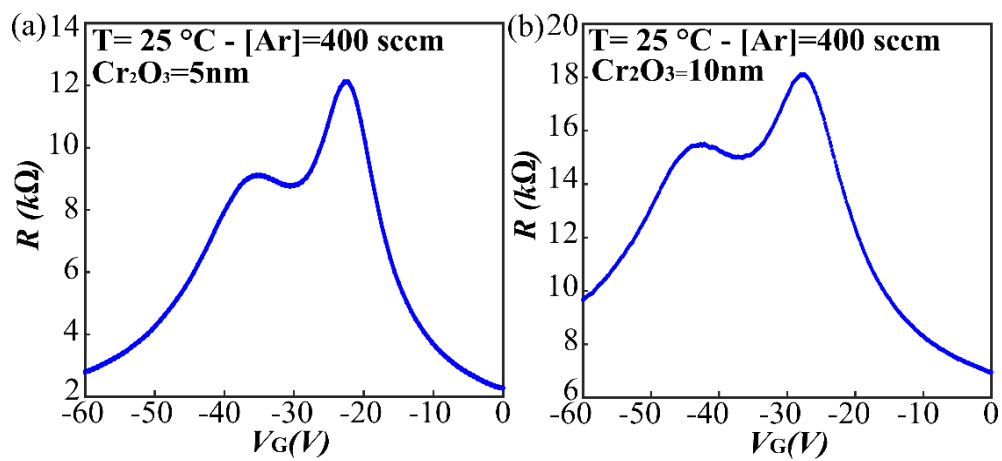

**Figure S8**. $RxV_G$ in Ar atmosphere (400 sccm) at $T$=25°C, for devices with 5 nm (a) and 10nm (b) of chromium oxide thick layer as stick layer for gold contacts.

Even though these devices exhibit double-CNP after fabrication in Argon atmosphere, when they are exposed to $H_2$ we observe a shift in both CNPs towards more negative gate value, indicating that both regions accept electrons (n-type doping) coming from the interaction with $H_2$ at the metal-graphene interface, as shown in Figure S9. One can note that initially the CNPs were positioned at $V_G^{ch} = -35V$ and $V_G^{cont} = -52V$ but after $H_2$ exposure the shifted to $V_G^{ch} = -43V$ and $V_G^{cont} < -60V$. These observations are also in agreement with our hypothesis that modifications at the metal-graphene interface can induce a large charge transfer to the graphene device.

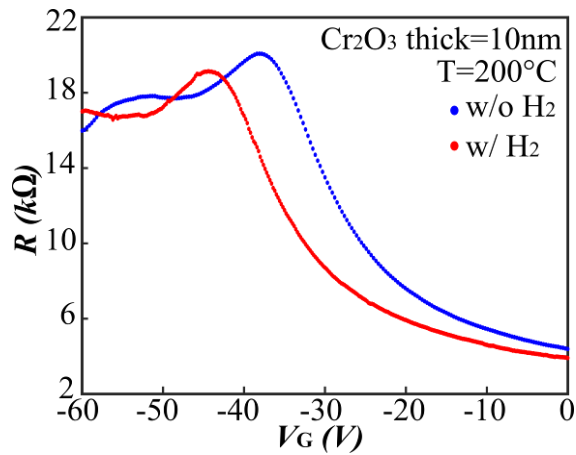

**Figure S9.** Two-probe $RxV_G$ for the graphene device with 10nm $Cr_2O_3$ thick at $T$=200°C. The blue curve presents the measurement without $H_2$, while the red curve shows the measurement with $[H_2]$=20% and after 1h of exposure.

**5 – Changes in the contact resistance ($R_C$) under $H_2$ exposure**

Figure S10 shows $RxV_G$ curves without (Fig. S10(a)) and with (Fig. S10(b)) $H_2$ exposure. In our previous work, we observed that hydrogen molecules are able to tune the doping underneath the contact, modulating the pn junction due to changes on electrostatic interaction between graphene and contact [6]. Here, we demonstrate this phenomenon clearly, as discussed in the main text and also in Figure S10. The curves present an increasing on resistance for negative values of charge density (*n*) more accentuated than the positive values of *n*, due the formation of second CNP.

In the main text we analyzed changes in the contact resistance ($R_C$) under $H_2$ exposure (Figure 2(d)) for a fixed value of *n*=-2.8x10$^{12}$cm$^{-2}$ (left side of the CNP in Figure

S10(b) marked by the dashed line). Now, we demonstrate that similar changes occur for values at $n= 1.45 \times 10^{12} cm^{-2}$ (right side of the CNP in Figure S10(b) marked by the dashed line), where the fitted $R_C$ values are $R_C^{Ar} = W.(460 \pm 80)\Omega\mu m$ and $R_C^{H2} = W.(430 \pm 20)\Omega\mu m$ for without and with $H_2$ exposure, respectively. Therefore, one can notice that the GFETs exposed to $H_2$ depict lower contact resistance, indicating that $R_C$ is affected by the molecules, $R_C^{H2}/R_C^{Ar} \approx 93\%$. These results suggest that the $H_2$ molecules can significantly change the metal-graphene interface potential, as expected and discussed in the main text.

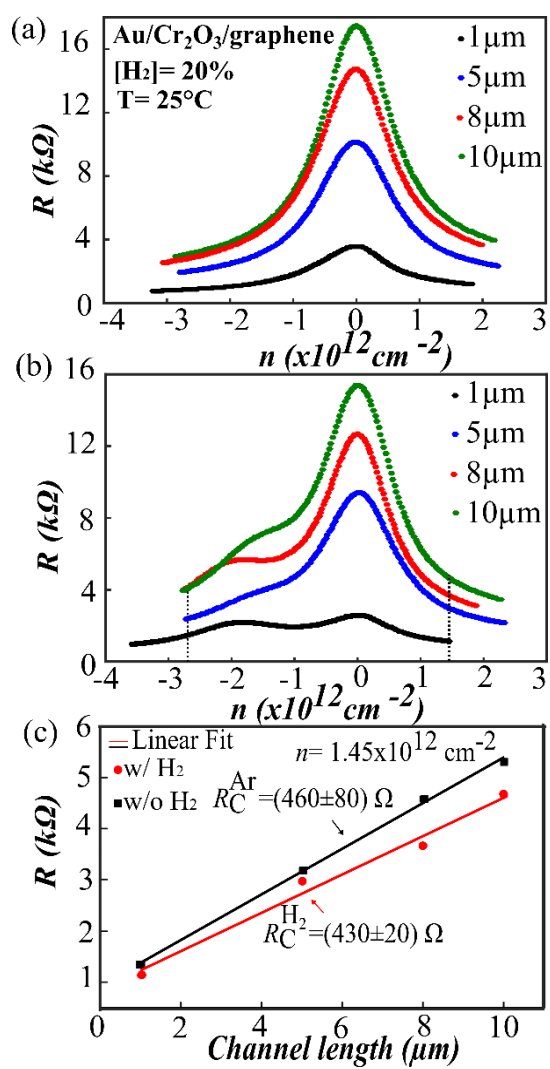

**Figure S8.** *RxV$_G$* for different channel length without normalization (a) without $H_2$ exposure and (b) with $H_2$ exposure. (c) *R x Channel length* w/ and w/o $H_2$ exposure.